# A complex chemical kinetic mechanism for the oxidation of gasoline surrogate fuels: n-heptane, iso-octane and toluene – Mechanism development and validation

**A. Pires da Cruz**[*,1], **C. Pera**[1], **J. Anderlohr**[1], **R. Bounaceur**[2] and **F. Battin-Leclerc**[2]

[1]IFP, France
[2]DCPR, UMR 7630 CNRS, INPL-ENSIC, France

**Abstract**
The development and validation against experimental results of a new gasoline surrogate complex kinetic mechanism is presented in this paper. The surrogate fuel is a ternary mixture of n-heptane, iso-octane and toluene. The full three components mechanism is based on existing n-heptane/iso-octane (gasoline PRF) and toluene mechanisms which were modified and coupled for the purpose of this work. Mechanism results are compared against available experimental data from the literature. Simulations with the PRF plus toluene mechanism show that its behavior is in agreement with experimental results for most of the tested settings. These include a wide variety of thermodynamic conditions and fuel proportions in experimental configurations such as HCCI engine experiments, rapid compression machines, a shock tube and a jet stirred reactor.

**Introduction**

Fundamental research in kinetic modeling of mixtures representative of commercial engine fuels (gasoline, diesel, jet...) has been strongly increasing in the last decade. Examples of such efforts can be found in references [1]-[9]. Motivation for this research has come from an increasing demand on understanding and modeling engine combustion towards a reduction in fuel consumption, $CO_2$ and pollutant emissions. However, commercial engine fuels may contain hundreds of different molecules and their composition is not completely well known. For modeling purposes, it is therefore necessary to define model fuels composed of molecules able to represent the most important classes of organic compounds present in real fuels as well as the most important features of the combustion process.

In this paper, the development and test of a kinetic mechanism for a surrogate fuel representative of gasoline used in spark ignition as well as gasoline Homogeneous Charge Compression Ignition (HCCI) internal combustion engines are presented.

Most past efforts in this area were concentrated in modeling gasoline PRF fuel mixtures composed of n-heptane and iso-octane in variable proportions. For this purpose, detailed mechanisms [2],[3] as well as reduced mechanisms have been developed [10],[11]. PRF mixtures are used for octane number determination and are representative of the spontaneous auto-ignition resistance in the conditions of the RON (Research Octane Number - procedure D-2699 [12]) and MON (Motor Octane Number - procedure D-2699 [12]) experiments. However, RON and MON are not representative of gasoline auto-ignition resistance in all possible engine conditions [13]. Also, laminar flame speeds of PRF mixtures are different from gasoline laminar flame speeds [14],[15] and may thus induce errors during flame propagation simulations.

Since kinetic modeling of large molecules includes hundreds of species involved in thousands of reactions, surrogate fuels cannot contain a large number of species in order to keep the chemical mechanisms manageable and computations possible. Gasoline is mostly composed of saturated hydrocarbons (normal and branched alkanes), olefins and aromatics. A three components gasoline surrogate including PRF mixtures and toluene has then been chosen here as a gasoline surrogate fuel. Toluene should allow better predictions of auto-ignition delays in various conditions since aromatics have a large octane number sensitivity (differences between RON and MON above 10) and different laminar flame speeds compared to alkanes. Olefins have not been considered in this work. The main reasons for this were: The surrogate mechanism described here is based on already well validated available individual component mechanisms and such a mechanism for olefins is still under development [16]; chemical behavior of unsaturated hydrocarbons is strongly dependant on the degree of unstauration making it difficult to choose one particular olefin; very few experimental results were found in the literature to allow for a thorough validation of a more complex surrogate mechanism. Olefin addition to the gasoline surrogate mechanism is however part of work in progress.

The full mechanism is described in the first part of the paper. Special attention is given to the coupling between the toluene and PRF sub-mechanisms. The second part of the paper deals with mechanism test and validation against a wide number of literature experimental configurations and thermodynamic conditions and surrogate fuel composition.

**Specific Objectives**

The target of the surrogate mechanism is to simulate the auto-ignition of three components mixtures with different percentages of each fuel in a variety of thermodynamic conditions representative of spark ignition and gasoline HCCI internal combustion engines. The proposed mechanism is validated against available experimental results from the literature. These include HCCI engine experiments, rapid compression machines and a shock tube. Jet stirred reactor simulations have also been compared to available experimental results. All simulations were performed

* Corresponding author: antonio.pires-da-cruz@ifp.fr, +33 1 47 52 65 02
1 et 4, Av. Bois Préau, 92852 Rueil Malmaison Cedex, France

with the Chemkin 4.1 combustion modeling package [17].

**Surrogate mechanism development**

The modeling approach followed in this paper is similar to that of Andrae et al. [4], who have also started from two different validated literature mechanisms: The PRF model from Lawrence Livermore National Laboratory [2] and the toluene mechanism from Dagaut et al. [18]. Validation of their mechanism was performed against Homogeneous Charge Compression Ignition (HCCI) engine experiments.

The surrogate mechanism developed is the present study is based on the available PRF mixtures (n-heptane and iso-octane) mechanism of Buda et al. [19] and on the toluene mechanism of Bounaceur et al. [20], with the addition of the needed co-oxidation reactions.

*PRF and toluene mechanisms*

The PRF mechanism of Buda et al. [19] and the toluene model of Bounaceur et al. [20] have the advantage of being issued from the same research group, meaning that they include a common $C_0$-$C_2$ reaction base. Thermochemical data of molecular and radical species have been calculated with the THERGAS software [21].

The PRF mechanism has been automatically generated by the EXGAS-ALKANE software [19] for the n-heptane/iso-octane mixture and includes all the usual low temperature reactions of alkanes (additions to oxygen, isomerization of peroxy radicals, formation of conjugated alkenes, cyclic ethers, aldehydes, ketones and hydroperoxides species).

The toluene model includes a sub-mechanism for the oxidation of benzene and of unsaturated $C_3$-$C_4$ species, as well as reactions of toluene and benzyl, tolyl (methylphenyl), peroxybenzyl, alcoxybenzyl and cresoxy free radicals in the primary mechanism and those of benzaldehyde, benzyl hydroperoxyde, cresol, benzylalcohol, ethylbenzene, styrene and bibenzyl in the secondary mechanism [20]. The simulations of experiments at pressures above 40 bar have required multiplying the rate constant of the decomposition of $H_2O_2$ by a factor 4 showing a not yet identified problem in the mechanism at high pressures.

These two mechanisms have already been validated in different experimental configurations under a wide range of thermodynamic conditions. They behave well in terms of auto-ignition delays measured in shock tubes and a rapid compression machine, although in some conditions, the agreement is not satisfactory for toluene. Mechanism results also agree well in terms of species evolution as a function of temperature measured in a jet stirred reactor.

*Co-oxidation reactions*

According to Andrae et al. [4], who have already proposed a coupled mechanism for n-heptane, iso-octane and toluene, co-oxidation reactions play an important role in both the auto-ignition of PRF mixtures as well as of n-heptane/toluene mixtures. These authors have proposed a set of coupling reactions which includes metathesis of primary fuels (RH) with R• radicals (heptyl, octyl and benzyl) and reactions between peroxy radicals (ROO•), toluene and PRF fuels.

Our previous work on PRF fuels has shown that co-oxidation reactions and species deriving from n-heptane and iso-octane were negligible. The coupling effect was only due to the pool of small radicals [22]. Concerning toluene, due to the strong electronic delocalization in benzyl radicals, co-oxidation involving these radicals has to be taken into account.

In summary, the following co-oxidation reactions have been written:

- Metathesis of benzyl radicals with n-heptane and iso-octane leading to toluene and alkyl radicals (respectively heptyl and octyl) with rate parameters given in Table 1;
- Terminations between benzyl and alkyl radicals, with a rate constant of $1.0 \times 10^{13}$ $cm^3.mol^{-1}.s^{-1}$ according to collision theory;
- Metathesis of secondary allylic radicals (iso-butenyl, iso-octenyl, heptenyl) with toluene with a rate constant of $1.6 \times 10^{12} \exp(-7600/T)$ $cm^3.mol^{-1}.s^{-1}$ as for allyl radicals [20];
- Reactions between toluene and alkyl peroxy radicals. A sensitivity analysis has shown that among all possible alkyl peroxy reactions, only methyl peroxy ($CH_3O_2$), due to its high concentration, needed to be considered, with a rate constant of $4.0 \times 10^{13} \exp(-6040/T)$ $cm^3.mol^{-1}.s^{-1}$ [4].

| Type of H-atom | A | n | $E_a$ |
|---|---|---|---|
| Primary | 1.6 | 3.3 | 19.84 |
| Secondary | 1.6 | 3.3 | 18.17 |
| Tertiary | 1.6 | 3.3 | 17.17 |

Table 1: Rate expressions for metathesis with benzyl radicals (in $s^{-1}$, kcal, mol units), taken from the values proposed by Tsang for allyl radicals [23].

**Results and Discussion**

The gasoline surrogate mechanism has been tested over a wide range of thermodynamic conditions for different surrogate fuel compositions. Table 2 summarizes the different gasoline three component fuel surrogates ran by each author and tested in this work. Experiments cover HCCI engine applications, rapid compression machines, a shock tube and a jet stirred reactor.

*HCCI engine, Andrae et al. [4]*

Andrae et al. [4] have performed experiments in a HCCI engine with PRF and n-heptane/toluene mixtures in the proportions given in Table 2 and operating conditions given in Table 3.

Simulations were ran from 99° Before Top Dead Center (BTDC) with initial pressure and temperature fit from experimental data. Since the auto-ignition delay calculation is the main objective of this work and since in an internal combustion engine, most heat losses occur after Top Dead Center, these have been neglected. The air/fuel mixtures were considered homogeneous.

| Authors | Setup | Fuel | iC8 | nC7 | Tol |
|---|---|---|---|---|---|
| Andrae et al. [4] | HCCI engine | A | 94 | 6 | - |
| | | B | 84 | 16 | - |
| | | C | - | 25 | 75 |
| | | D | - | 35 | 65 |
| Tanaka et al. [24] | RCM | A | - | 26 | 74 |
| | | B | 5 | 21 | 74 |
| | | C | 10 | 16 | 74 |
| Vanhove et al. [25] | RCM | A | - | 50 | 50 |
| | | B | 65 | - | 35 |
| Gauthier et al. [26] | ST | A | 55 | 17 | 28 |
| | | B | 63 | 17 | 20 |
| Dubreuil et al. [27] | JSR | A | - | 80 | 20 |

Table 2: Three components fuel mixtures tested in the different authors experiments and used in this work.

| Op. Cond. | RPM | $T_{init}$ (°C) | $p_{init}$ (bar) | $\phi$ |
|---|---|---|---|---|
| Run 1 | 900 | 120 | 1 | 0.29 |
| Run 2 | 900 | 40 | 2 | 0.25 |

Table 3: HCCI engine operating conditions of Andrae et al. [4]. Compression ratio: 16.7.

Figure 1 and Figure 2 show that for both PRF fuel and n-heptane/toluene mixtures, the mechanism reproduces the sensitivity of the ignition delay to the fuel composition and the initial engine conditions. Over-prediction of the maximum pressure is due to both adiabatic and homogeneous simulations whereas in the engine, the load is never really homogeneous and heat losses play an important role during piston expansion.

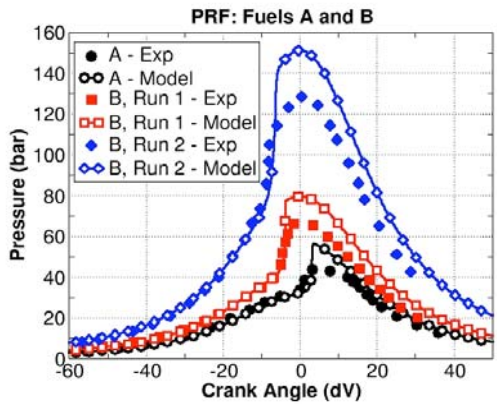

Figure 1: Experimental and simulated pressure curves for fuels A (Run 1) and B (Run 1 and Run 2). Initial temperature and pressure at start of simulations (99° BTDC) – Fuel A/Run 1: 460 K, 1.2 bar, Fuel B/Run 1: 470 K, 1.7 bar, Fuel B/Run 2: 420 K, 3.2 bar. HCCI engine of Andrae et al. [4].

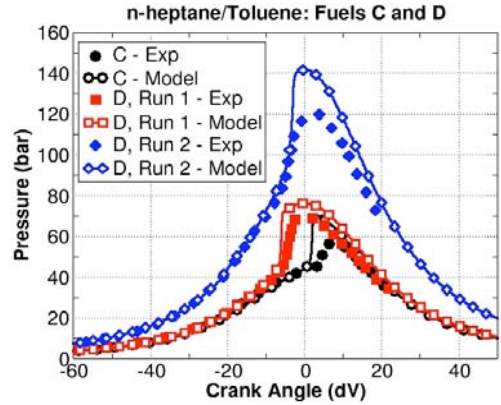

Figure 2: Experimental and simulated pressure curves for fuels C (Run 1) and D (Run 1 and Run 2). Initial temperature and pressure at start of simulations (99° BTDC) – Fuel C/Run 1: 470 K, 1.6 bar, Fuel D/Run 1: 470 K, 1.6 bar, Fuel D/Run 2: 420 K, 3.1 bar. HCCI engine of Andrae et al. [4].

*Rapid Compression Machine, Tanaka et al. [24]*

Tanaka et al. [24] have studied the auto-ignition behavior of different fuels in a Rapid Compression Machine (RCM) over a variety of thermodynamic conditions. We have chosen here a fixed temperature and pressure (318 K and 0.1 MPa respectively) and three n-heptane, iso-octane and toluene mixtures (cf. Table 2) in order to assert the sensitivity of the mechanism to small variations of the iso-octane to n-heptane ratio with a fixed amount of toluene. Results presented in Figure 3 show this sensitivity is well captured by the mechanism although the simulated auto-ignition delay differences between fuels are smaller than the experimental ones.

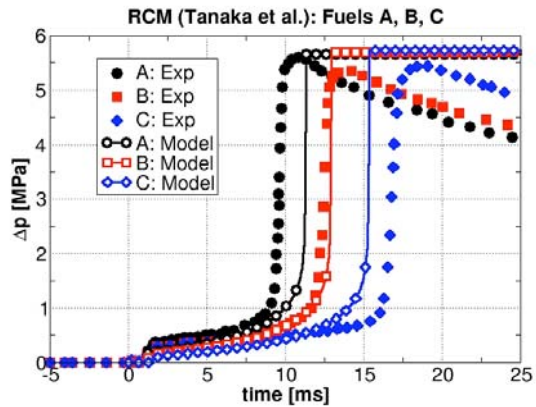

Figure 3: Experimental and simulated pressure change (Δp) curves for fuels A, B and C. Equivalence ratio: 0.4, Initial temperature: 318 K, Initial pressure: 0.1 MPa, Compression ratio: 16. Rapid Compression Machine of Tanaka et al. [24].

*Rapid Compression Machine, Vanhove et al. [25]*

Vanhove et al. [25] have studied the auto-ignition as a function of temperature of a 50/50 (volume) n-heptane and toluene mixture and a 65/35 iso-octane and toluene mixture in another RCM. They have varied the initial temperatures and pressures and measured both cool flame ($t_{cf}$) and main ignition delays ($t_{ig}$). As a reference, auto-ignition delays of pure n-heptane and pure iso-octane have been measured in similar thermodynamic conditions and simulated with the same kinetic mechanism.

Figure 4 shows that both ignition delays are well captured by the mechanism and that the retarding influence of toluene over n-heptane is retrieved. Nevertheless, for the latter case, the Negative Temperature Coefficient (NTC) zone is not well reproduced. The mechanism does not show enough reactivity between lower and intermediate temperatures.

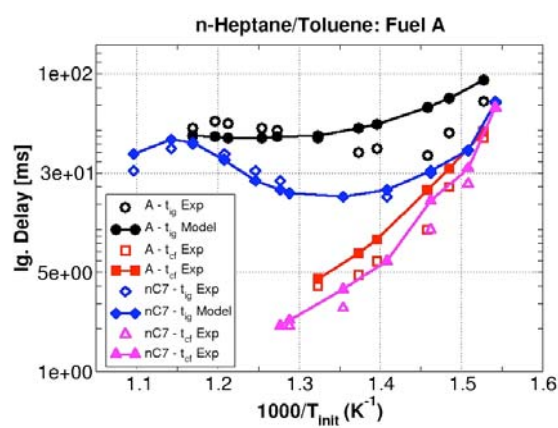

Figure 4: Experimental and simulated main ($t_{ig}$) and cool flame ($t_{cf}$) auto-ignition delays for fuel A and n-heptane. Pressures between 3.7 bar and 4.9 bar and Equivalence ratio: 1. Rapid Compression Machine of Vanhove et al. [25].

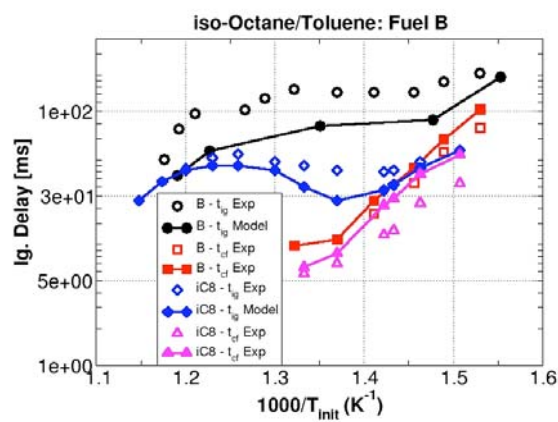

Figure 5: Experimental and simulated main ($t_{ig}$) and cool flame ($t_{cf}$) auto-ignition delays for fuel B and iso-octane for pressures between 12 bar and 16.1 bar and Equivalence ratio: 1. Rapid Compression Machine of Vanhove et al. [25].

Figure 5 represents the same ignition data in the case of the iso-octane/toluene and pure iso-octane mixtures. Again, the main observed experimental trends are captured by the mechanism. However, the strong inhibiting effect of toluene on iso-octane observed by Vanhove et al. is underestimated by the mechanism. The authors have attributed this effect to a deactivation of the radical pool caused by the presence of toluene. Toluene can react either with OH or $HO_2$ to form unreactive species such as benzyl, $H_2O$ or $H_2O_2$ respectively. This has been confirmed by studying the sensitivity of the mechanism to the kinetic constant of reaction toluene+OH leading to benzyl+$H_2O$ which was shown to be very high. The interactions between iso-octane and toluene at the radical pool level are thus an open subject and are the subject of work in progress.

*Shock tube, Gauthier et al. [26]*

Gauthier et al. [26] have studied the ignition of n-heptane, iso-octane and toluene mixtures at high temperature and pressure in a shock tube. Their experiences also covered a broad range of equivalence ratios and dilution gas proportions. We have chosen to present their results at stoichiometry and zero dilution although all reported experimental conditions have been tested.

Figure 6 shows that the mechanism is again able to capture the influence of coupled temperature, pressure and fuel differences in terms of auto-ignition delays in conditions very close to those inside an internal combustion engine. Results of simulations at different equivalence ratios and dilution gas concentrations have also shown to agree well with experimental results.

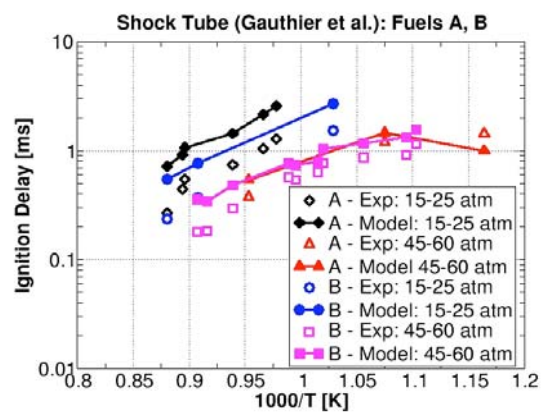

Figure 6: Experimental and simulated auto-ignition delays for surrogate fuels A and B for pressures between 12 atm and 60 atm, Equivalence ratio: 1 and initial temperatures between 860 K and 1022 K. Shock tube of Gauthier et al. [26].

*Jet stirred reactor, Dubreuil et al. [27]*

In spite of the fact that our first objective is to conveniently simulate auto-ignition delays of representative engine fuels, it is also interesting to test the species evolution predicted by the mechanism against available Jet Stirred Reactor (JSR) results. In this context, we have simulated the Dubreuil et al. [27] n-heptane/toluene mixture experiment where the authors have measured reactant as well as main intermediate species molar concentrations as a function of temperature.

Figure 7 shows that both n-heptane and toluene evolutions as a function of temperature are well captured by the mechanism although around 650 K to 700 K, the mechanism is not reactive enough with a slower consumption of n-heptane compared to experiments. This agrees with the discrepancies obtained in the same temperature range by Vanhove et al. [25], as it can be seen in Figure 5.

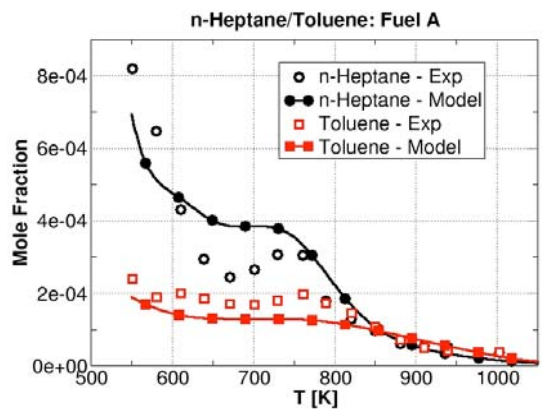

Figure 7: Experimental and simulated mole fractions of n-heptane and toluene as of function of temperature for fuel A. Pressure: 10 bar, Equivalence ratio: 0.2, Residence time=0.5 s. Jet Stirred Reactor of Dubreil et al. [27].

It is however interesting to note that one main combustion product ($CO_2$) as well as two main intermediates (CO and formaldehyde) evolutions as a function of temperature are well reproduced by the mechanism (see Figure 8).

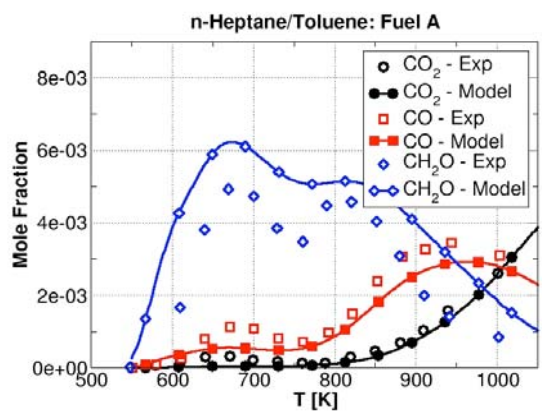

Figure 8: Experimental and simulated mole fractions of $CO_2$, CO and $CH_2O$ as of function of temperature for fuel A. Pressure: 10 bar, Equivalence ratio: 0.2, Residence time=0.5 s. Jet Stirred Reactor of Dubreil et al. [27].

**Conclusions**

A three components chemical kinetic mechanism developed to model gasoline surrogate fuel ignition and combustion was developed. The full mechanism is based on existing n-heptane/iso-octane and toluene sub-mechanisms. The main objective of this paper was to present the coupling strategy between the sub-mechanisms and to test the final version against experimental results available from the literature.

Simulations with the three components gasoline surrogate mechanism show that its behavior is in agreement with experimental results for most of the conditions tested. We emphasize the model predictivity in terms of simulating auto-ignition delays of different fuel mixtures for different temperatures, pressures and equivalence ratios covering most internal combustion engines operating conditions as well as different experimental setups.

The auto-ignition delays measured by Andrae et al. [4] in a HCCI engine with n-heptane/iso-octane and n-heptane/toluene mixtures are reproduced by the mechanism. The simulations of the Tanaka et al. [24] experiments in a rapid compression machine show that the mechanism is also capable of retrieving the sensitivity of auto-ignition delay variations to very small iso-octane addition to toluene and n-heptane mixtures. The major trends of the results obtained by Vanhove et al. [25] in a rapid compression machine have been captured as well as the Gauthier et al. [26] shock tube auto-ignition measurements with two different three components surrogate fuels. Finally, the simulation of the Dubreuil et al. [27] experiments in a jet stirred reactor shows that major species evolutions as a function of temperature of a binary toluene/n-heptane mixture are also conveniently captured by the three components mechanism.

Progress topics have been identified and are the subject of work in progress. These include a better coupling between n-heptane and toluene and iso-octane and toluene respectively in the NTC region and on the radical pool level and also the development of a gasoline surrogate including olefins, a major component of this fuel.

**Acknowledgements**

This research was partially funded by the GSM (Groupement Scientifique Moteur: PSA Peugeot Citroen, Renault and IFP).